\documentclass[%
 reprint,
superscriptaddress,
 amsmath,amssymb,
 aps,
]{revtex4-2}

\usepackage[utf8]{inputenc}
\usepackage{braket}

\usepackage{graphicx}
\usepackage{dcolumn}
\usepackage{bm}
\usepackage{hyperref}
\usepackage{balance}
\usepackage{mathrsfs}
\usepackage{todonotes}

\begin{document}

\title{Graph Neural Network Autoencoders for Efficient Quantum Circuit Optimisation}

\author{Ioana Moflic}
\email{ioana.moflic@aalto.fi}
\affiliation{Department of Computer Science, Aalto University, 00076 Espoo, Finland}

\author{Vikas Garg}
\email{vgarg@csail.mit.edu}
\affiliation{Department of Computer Science, Aalto University, 00076 Espoo, Finland}
\affiliation{YaiYai Ltd}

\author{Alexandru Paler}
\email{alexandru.paler@aalto.fi}
\affiliation{Department of Computer Science, Aalto University, 00076 Espoo, Finland}

\begin{abstract}
Reinforcement learning (RL) is a promising method for quantum circuit optimisation. However, the state space that has to be explored by an RL agent is extremely large when considering all the possibilities in which a quantum circuit can be transformed through local rewrite operations. This state space explosion slows down the learning of RL-based optimisation strategies. We present for the first time how to use graph neural network (GNN) autoencoders for the optimisation of quantum circuits. We construct directed acyclic graphs from the quantum circuits, encode the graphs and use the encodings to represent RL states. We illustrate our proof of concept implementation on Bernstein-Vazirani circuits and, from preliminary results, we conclude that our autoencoder approach: a) maintains the optimality of the original RL method; b) reduces by 20 \% the size of the table that encodes the learned optimisation strategy. Our method is the first realistic first step towards very large scale RL quantum circuit optimisation.
\end{abstract}

\maketitle

\section{Introduction}

Scalable optimisation methods for quantum circuits are an open problem. The NISQ generation of quantum computers, even when operating on thousands of qubits, will not be fully error-corrected such that structural properties of the compiled circuits (e.g. depth, number of gates) play a major role in estimating the failure-rate of the entire computation. In particular, deeper circuits have a higher failure rate. Without using quantum error-correction or error-mitigation, one should compile aggressively the circuits to reduce their depth, and then use error mitigation methods (e.g. \cite{endo2018practical, smith2021error}).

Consequently, compilation scalability is necessary for achieving the fault-tolerant execution of the first large scale quantum computation. The technical road maps project that quantum computers will operate thousands of qubits within the next few years, but there is a gap between the speed of the current generation of quantum circuit compilers and the speed required for circuits operating on thousands of qubits.

\subsection{Motivation}

Machine learning techniques have started being applied to quantum circuit compilation and optimisation (e.g. \cite{paler2020machine, zulehner2019evaluating, krenn2023artificial}). The general approach is to invest large amounts of computational power into the training of models that can then be used for fast and efficient quantum circuit compilation.

Machine learning techniques are increasingly successfully applied for classical circuit design automation\cite{gubbi2022survey}. Reinforcement Learning (RL) for the compilation of quantum gate sequences is presented in~\cite{moro2021quantum}. At very large scale, RL has been successfully applied for classical chip design~\cite{mirhoseini2021graph}. Small scale compilation of quantum circuits is shown in ~\cite{moro2021quantum, fosel2021quantum} and the potential of RL for quantum circuit compiler optimisation has been illustrated by~\cite{quetschlich2022compiler}. Large scale applications of RL with respect to quantum circuits have not been demonstrated by now, and our work is a first step towards scalable quantum circuit optimisation.

\begin{figure}[!t]
    \centering
    \includegraphics[width=0.95\columnwidth]{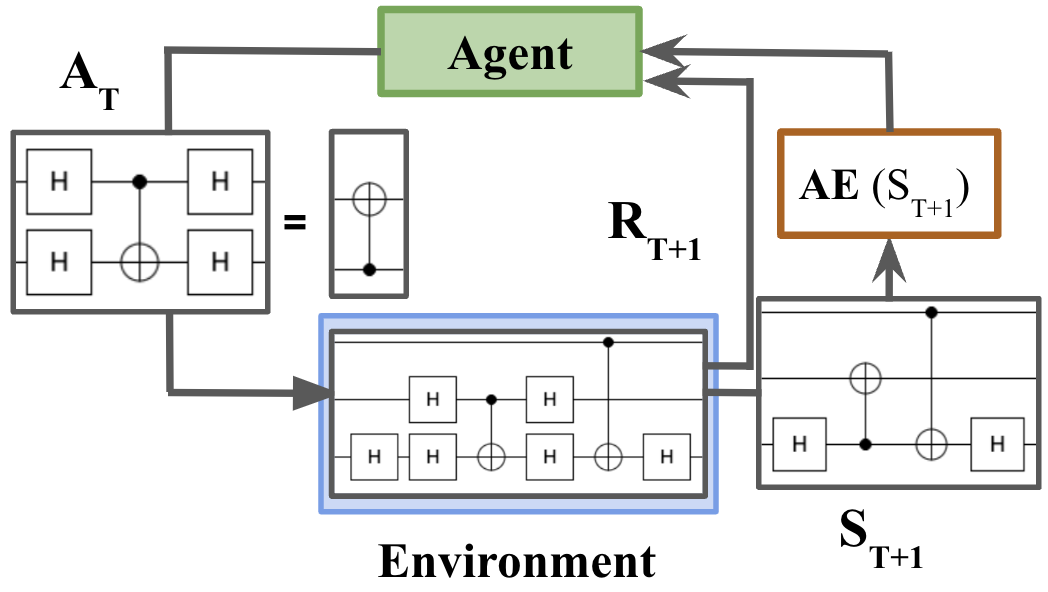}
    \caption{The learning loop of the RL agent with embedded autoencoder for the representation of the environment observation. At each step of the RL algorithm, the agent chooses an action $A_{T}$ to apply and is given the reward $R_{T+1}$ and the encoding $AE(S_{T+1})$ of the transformed circuit.}
    \label{fig:architecture}
\end{figure}

This paper is organised as follows: Section~\ref{sec:opt}--\ref{sec:rl} are introducing the background necessary for presenting the method which we detail in Section~\ref{sec:met}. Therein, we focus on the data structure necessary for training the autoencoder that is afterwards embedded into RL. Finally, we present preliminary results collected by evaluation our implementation on benchmarking circuits.

\subsection{Quantum Circuit Optimisation}
\label{sec:opt}

\begin{figure}[t]
    \centering
    \includegraphics[width=0.8\columnwidth]{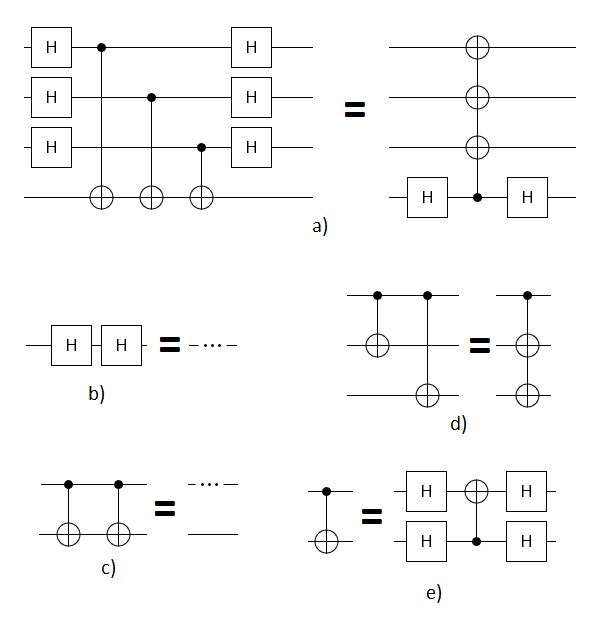}
    \caption{a) Unoptimized and optimized Bernstein-Vazirani circuit; b) Two Hadamard gates cancelling; c) Two CNOT gates cancelling; d) Parallelizing CNOTs sharing the same control qubit; e) reversing the direction of a CNOT using Hadamard gates.}
    \label{fig:circs}
\end{figure}

Quantum circuit optimisation using template-based rewrite rules \cite{miller2003transformation} is widely used in quantum circuit software (e.g. Google Cirq~\cite{cirq}, IBM Qiskit~\cite{qiskit}). An input circuit is gradually transformed by applying quantum gate identities (Fig.~\ref{fig:circs}) until a given optimisation criterion is met. The gate set and the size of the input circuits influence the performance of the procedure. The number of permitted transformations blows up the size of the optimisation search space. Consequently, although this kind of optimisation performs well, it is challenging to improve its scaling.

\subsection{Reinforcement Learning}
\label{sec:rl}

Quantum circuit optimisation can be framed as a RL problem. Given the quantum circuit and a range of templates to be applied, an agent would learn an optimal policy in a trial and error approach. The circuit represents a fully observable environment and applicable templates at a given time step are the actions. The agent is selecting actions from an action space, which is formed of circuit rewrite rules (templates) as illustrated in Fig.~\ref{fig:circs}. Each action transforms a quantum circuit into a functionally equivalent, but structurally different quantum circuit. The structure of the circuit at a given time step is expressed by an observation of the environment.

The state space in RL is formed by all the states encountered by the agent during training and all the actions that allowed the agent to transition between those states form the action space. In canonical RL, the mapping between states and actions is stored in a table, also called Q-Table, whereas in deep RL the mapping is learnt by a machine learning model.

The Q-Table is effectively the encoding of the optimisation algorithm that the agent learned during training. The size of the Q-Table increases faster at the beginning of learning, when the agent is \emph{exploring} the environment. The Q-Table's size is increasing slower towards the end of the learning: the agent will \emph{exploit} the knowledge it accumulated.

\subsection{Graph Neural Networks}

Graph neural networks (GNN)~\cite{gnn, gnn_2} are a type of neural networks capable of performing machine learning tasks on graph-structured data. GNNs rely on message passing between the nodes of the graph, where messages which are in the form of vectors are updated using neural networks. At each message passing step, nodes aggregate the information received from the other nodes in proximity. This type of neural network is commonly used for link prediction~\cite{link_pred}, community detection~\cite{gnn_comm}, graph classification~\cite{gnn_class}, etc. Graph Autoencoders are autoencoders which use GNNs for the encoder and the decoder components of an autoencoder.

Variational autoencoders for directed acyclic graphs (D-VAE)~\cite{zhang2019d}, are effective in finding an encoding for DAGs into the latent space of the autoencoder. Grammar Variational Autoencoders~\cite{kusner2017grammar}, as an example of D-VAEs, encode and decode molecules into and from a continuous space, and by searching in that specific space, valid, optimised molecule forms can be found. D-VAEs are also useful for learning distributions of approximate, lossy circuit representations. In the context of RL, the goal is to find a lossy state representation of the RL environment that is precise enough such that the RL-agent can differentiate between the consequences of its actions and 
at the same time, sufficiently imprecise in order to be meaningful for more than one RL-state.

\subsection{Contributions}

This work is the first to present GNN autoencoders~\cite{zhang2019d} for speeding up the RL optimisation of quantum circuits. We achieve a more efficient (compact) RL optimisation compressing the Q-Table using a GNN autoencoder~\cite{kingma2019introduction}. 

We use autoencoders to describe quantum circuits using a probabilistic encoding. Instead of building a deterministic function which outputs a unique encoding for each circuit, we are using the encoder to describe a probability distribution for each snapshot of the RL environment's state. encoded representations.

Embedding autoencoders into the RL procedure is a recent method \cite{lange2010deep, prakash2019use} which has not been explored, to the best of our knowledge, for quantum circuit optimisation. Our approach is promising in the context of RL-optimisation of quantum circuits: a) dimensionality reduction of the Q-table~\cite{Sutton1998}; b) improving convergence time by finding an efficient encoding while maintaining a good performance after decoding.

We have implemented our method (Section~\ref{sec:res}) and obtained empirical evidence about the scalability and efficiency of our method. We demonstrate scalability by training our RL agent on Bernstein-Vazirani circuits operating on 2-5 qubits.

\section{Methods}
\label{sec:met}

We present the methods we used to implement the workflow illustrated in Fig.~\ref{fig:architecture}. Without loss of generality, we restrict the quantum circuits we are optimising to ICM+H circuits. ICM circuits~\cite{paler2017fault} are related to measurement-based graph-state quantum circuits~\cite{vijayan2022compilation}. ICM circuits are computationally universal and consist of single qubit initialisation, CNOT gates, single qubit measurements. In order to implement the correct computation, ICM circuits will also rely on classical feedback. ICM+H circuits are ICM circuits which include single qubit Hadamard gates, too.

Herein we use a limited number of templates, circuit rewrite rules, as illustrated in Fig.~\ref{fig:circs}b-e). For benchmarking purposes (see the Results section), we will use Bernstein-Vazirani (BV) circuits (Fig.~\ref{fig:circs}a). The BV circuits have practically the ICM+H form. BV circuits have a known optimum depth of three, which can be achieved if the templates from Fig.~\ref{fig:circs} are used in an optimal order.

In the following, we discuss how our RL framework is operating, how to obtain lossy circuit representations with an autoencoder, and the method to include the autoencoder into the training of RL.

\subsection{Reinforcement Learning with a Lossy Representation}

Our RL method is based on Q-Learning, which is a model-free algorithm~\cite{watkins1989learning}. The RL environment is explored by an agent, which chooses an action (circuit template) to apply at each step. There is a reward associated with each action, and the value of the reward is reflecting the environment's response to the action. We use relatively small learning environments (compared to computer games where reinforcement learning has been successfully applied), but there still exists a combinatorial explosion in the number of possibilities how the templates can be applied.

Each of the agent's actions is transforming the structure of the circuit. Assuming that $g_i$ is the circuit before applying the template $t$, and that $g_o$ is the circuit afterwards, we can define the states $\mathcal{s}(g_i)$ and $\mathcal{s}(g_o)$. The RL agent will encode in the QTable the transition $\mathcal{s}(g_i) \xrightarrow[]{\text{t}} \mathcal{s}(g_o)$. 

The $\mathcal{s}$ function is used for identifying circuits. For example, $\mathcal{s}_s$ might be the character string representation of the circuit's QASM gate list (e.g. for two CNOTs in the circuit ``\texttt{cx 0 1, cx 1 0}''). The character string representation can identify each encountered circuit uniquely, and this can be a disadvantage because of the very large number of encountered states and the size of the resulting QTable. States are added to the table if these did not exist beforehand. For example, assuming that $\mathcal{s}(g_i)$ did exist, and that $\mathcal{s}(g_o)$ is novel, then the number of encountered states in the table increases by one -- and this can be the case after each application of a template.

We limit the growth of the QTable size by using lossy versions of $\mathcal{s}$. A lossy representation may seem inefficient, but it has the advantage of abstracting circuits into classes, where a class denotes a set of circuits with similar structure. This is a powerful property because a specific optimisation template is very likely to be beneficial for similar circuits. For example, assuming that $g_i$ and $g_o$ have the same meaning as in the example above, but that $\mathcal{s}(g_i) == \mathcal{s}(g_o)$, then the number of encountered states would not increase. This is only to say that a particular template $t$ leaves, from the perspective of $\mathcal{s}$, the circuits unchanged: $g_i$ and $g_o$ belong to the same circuit class.

\subsection{Formalising DAGs for the Encoder}

\begin{figure}[!t]
    \centering
    \includegraphics[width=0.95\columnwidth]{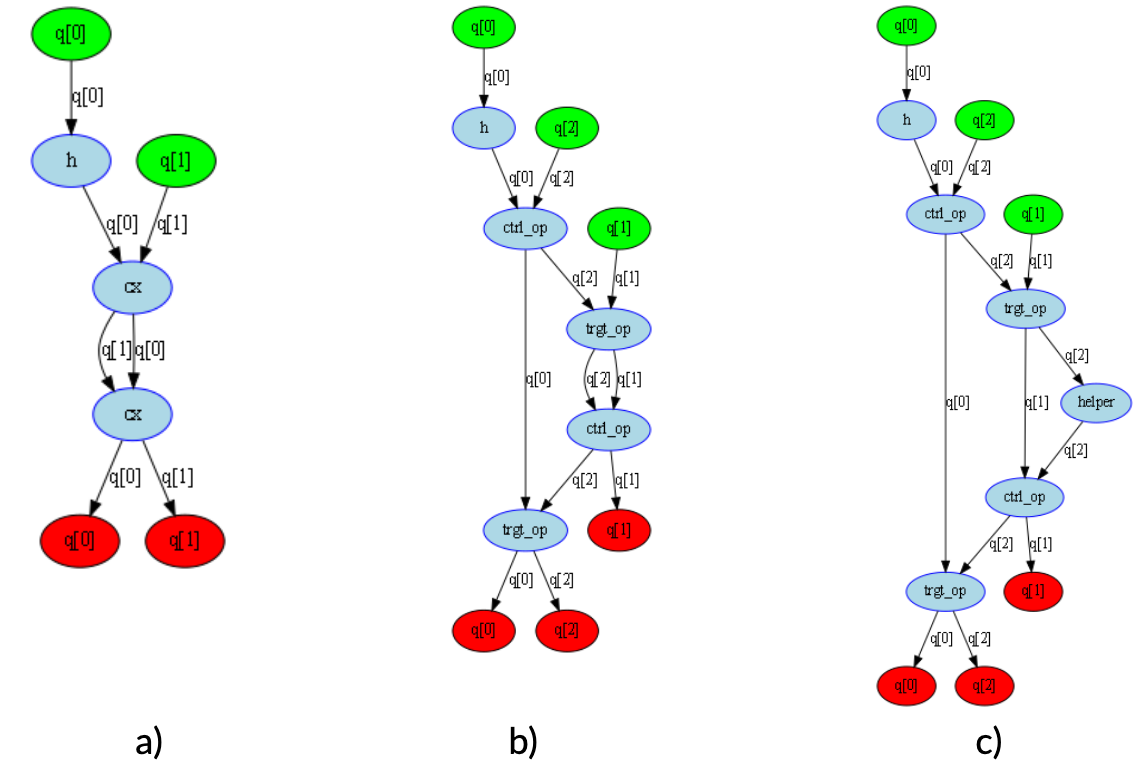}
    \caption{Every quantum circuit is a DAG, but not every DAG is a valid quantum circuit. Issues arise when sequences of CNOT gates have to be represented as DAGs: there have to be different graph nodes for control and target, or to annotate the edges. We choose the first option. a) A DAG where two CNOTS (\texttt{cx}) are applied on the same qubits (\texttt{q[0]} and \texttt{q[1]}). b) Introducing two distinct node types (\texttt{trgt\_op} and \texttt{ctrl\_op}) in order to differentiate between the CNOT gate orientations. In order to maintain the property that each quantum circuit gate operate has an equal number of input and output qubits, the DAG nodes have to have an equal number of predecessor and successor nodes -- we introduce a \emph{fake wire} (\texttt{q[2]}) that is connecting the control and target nodes of the same CNOT gate. c) An additional node type is used (\texttt{helper}) in order to make clear which wire is the \emph{fake} when a target and a control node operate on the same pair of wires.}
    \label{fig:mydag}
\end{figure}

Quantum circuits can be represented as directed acyclic graphs (DAGs). The goal is to treat each quantum circuit as a data point and to encode the point into a latent distribution of the autoencoder (see next section). To this end, we use graph neural networks (GNN)~\cite{gnn} which are fed with formally correct DAG representations of quantum circuits. Fig.~\ref{fig:mydag} illustrates the three additional node types (\texttt{trgt\_op}, \texttt{ctrl\_op}, \texttt{helper}) we use in order to have DAGs which are correct from the perspective of quantum circuit representations.

\subsection{Training the Autoencoder}

DAGs have a dependency structure, and one can analyse them as a single computation formed by the topological sorting of the nodes. Our D-VAE employs an asynchronous message passing scheme to injectively encode the DAG's computation~\cite{zhang2019d}. For encoding, we use the information about the DAG's node types (\texttt{input, output, control, target, hadamard, helper}), and the edges existing between the nodes.

For each node of the quantum circuit's DAG, we use a Gated Recurrrent Unit (GRU)~\cite{chung2015gated} to compute a corresponding hidden state. The latter is a function of the hidden states  collected from the node's neighbours. The encoding process guarantees two properties~\cite{zhang2019d}: 1) two isomorphic DAGs have the same encoding; b) encoding does not differentiate between two circuits $g_1$ and $g_2$ as long as they represent the same computation. The D-VAE finds a single vector encoding for the two similar graphs. We are using this property for representing similar RL states. 

The decoder uses the same asynchronous message passing scheme as the encoder, but in reverse. The decoder reconstructs a generated DAG node by node, after sampling a node type distribution and an edge probability distribution from the autoencoder's latent space.

The D-VAE is trained by backpropagating through the GRU's parameters the partial derivatives of a loss function $\mathcal{L}$. In the equation below, $\alpha$ and $\gamma$ are scaling factors, $g$ is the original DAG that is encoded, $g'$ is the decoded DAG, $\mathcal{R}(g,g')$ is the reconstruction error function which uses binary cross-entropy between the node type and edge probability distributions of $g$ and $g'$, and $\mathcal{E}(e, g')$ is the edge edit distance between $g$ and $g'$ :
\begin{align*}
    \mathcal{L}(g, g') =\alpha \mathcal{R}(g,g') + \gamma \mathcal{E}(g, g')
\end{align*}

\section{Results}
\label{sec:res}

\begin{figure}[!t]
\centering
\includegraphics[width=0.99\columnwidth]{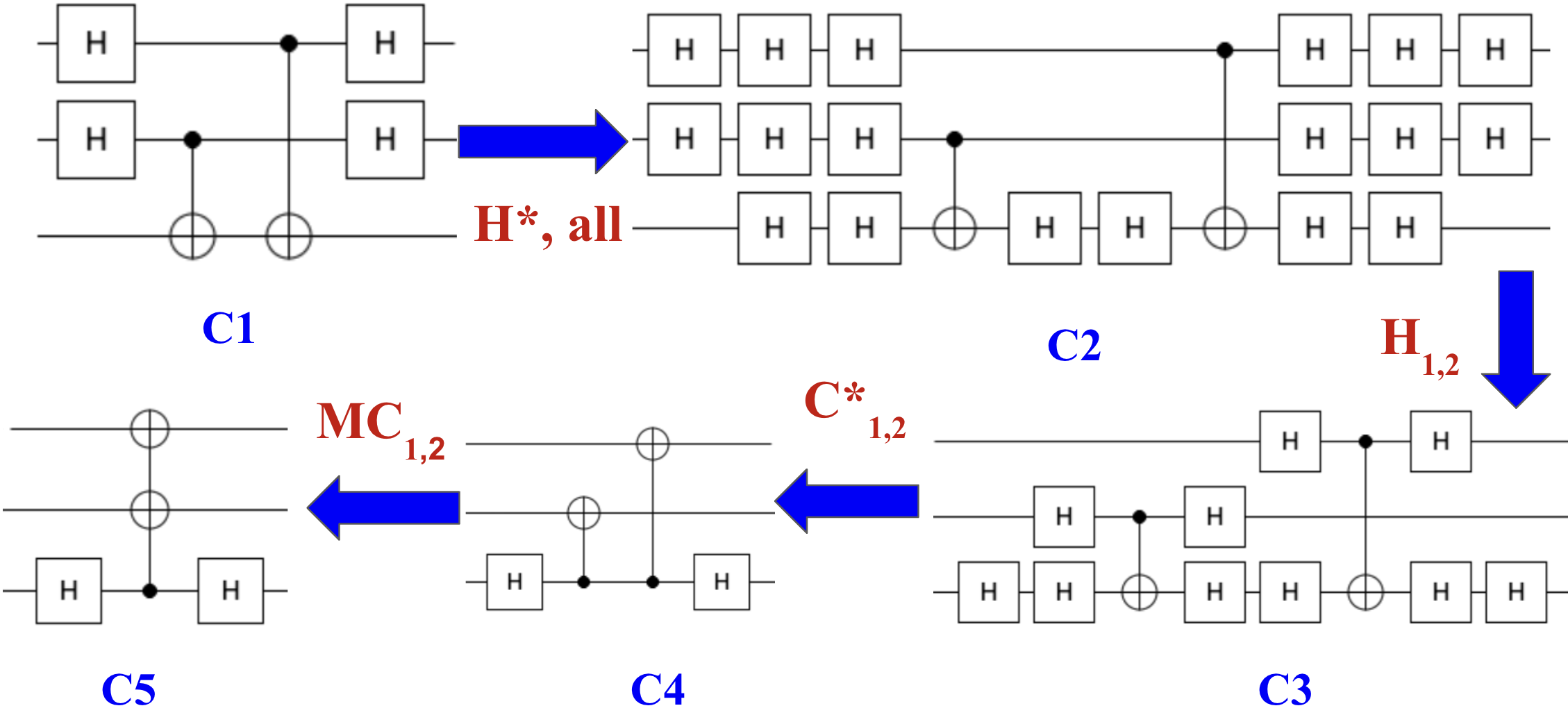}
\caption{Example of how a 2-qubit Bernstein Vazirani circuit might be optimized. $C_i$ are the intermediate states of the circuit obtained by applying the templates from Fig.\ref{fig:circs}. The transitions annotated on the arrows, indicate if a template was applied directly (e.g. \texttt{H}, for cancelling two Hadamards), or in reverse (e.g. \texttt{H*}, for inserting two Hadamards next to each other). A template is applied on a set of qubits (e.g. \texttt{1,2}) or on all qubits (\texttt{all}). Depending on the choice of the function $\mathcal{s}$, this sequence will be encoded into different QTables -- two possibilities are illustrated in Fig.\ref{fig:mdps}. }
\label{fig:example}
\end{figure}

We implemented our method using Google Cirq~\cite{cirq} (to apply the template rewrite rules), IBM Qiskit~\cite{qiskit} (to convert a quantum circuit to a DAG), OpenAI Gym\cite{openai} (the RL engine) and a custom variant of the variational autoencoder presented in~\cite{zhang2019d}. The autoencoder is used within the RL workflow to compute the representation of the encountered RL states.

\begin{figure}
    \centering
    \includegraphics[width=0.95\columnwidth]{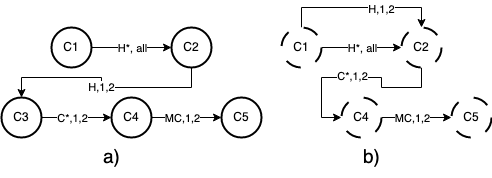}
    \caption{Two different RL state transitions obtained by using different state encodings for the circuit transformations from Fig.\ref{fig:example}. a) using an exact character string representation there are four states; b) a lossy encoding, such as the one from the autoencoder, might determine that the representations of C2 and C3 are the same, and the that there are two possibilities from starting from C1 into any of those two states.}
    \label{fig:mdps}
\end{figure}

We compare the effectiveness and speed of our method with a vanilla RL Q-Learning implementation, and use the following benchmarking  procedure. We train an RL agent, called $R_s$, with a non-lossy circuit representation (e.g. character string). After the training we collect all the states from the QTable and use these for training an GNN autoencoder. Let $l_s$ be the number of states from this QTable. We repeat the training of an RL agent, called $R_a$ from scratch, but this time using the autoencoder and obtain an $l_a$ number of states in the QTable. We consider that the autoencoder resulted in a more \emph{effective} QTable representation if $l_a < l_s$.

We used Bernstein-Vazirani circuits for benchmarking purposes. These circuits are advantageous because they have a known global optimum: when all the CNOTs are parallelised and the total depth is three. The goal of the RL method is to learn the strategy that generates the minimum depth circuit after parallelising the CNOT gates (Fig.~\ref{fig:circs}a). This type of CNOT gate parallelism is compatible with surface code error-corrected quantum circuits~\cite{fowler2012surface} implemented by braiding or lattice surgery. In order to build an intuition of the our results, Fig.~\ref{fig:example} illustrates the application of our method for the optimisation of a simple circuit.

\begin{figure}[!t]
    \centering
    \includegraphics[width=0.95\columnwidth]{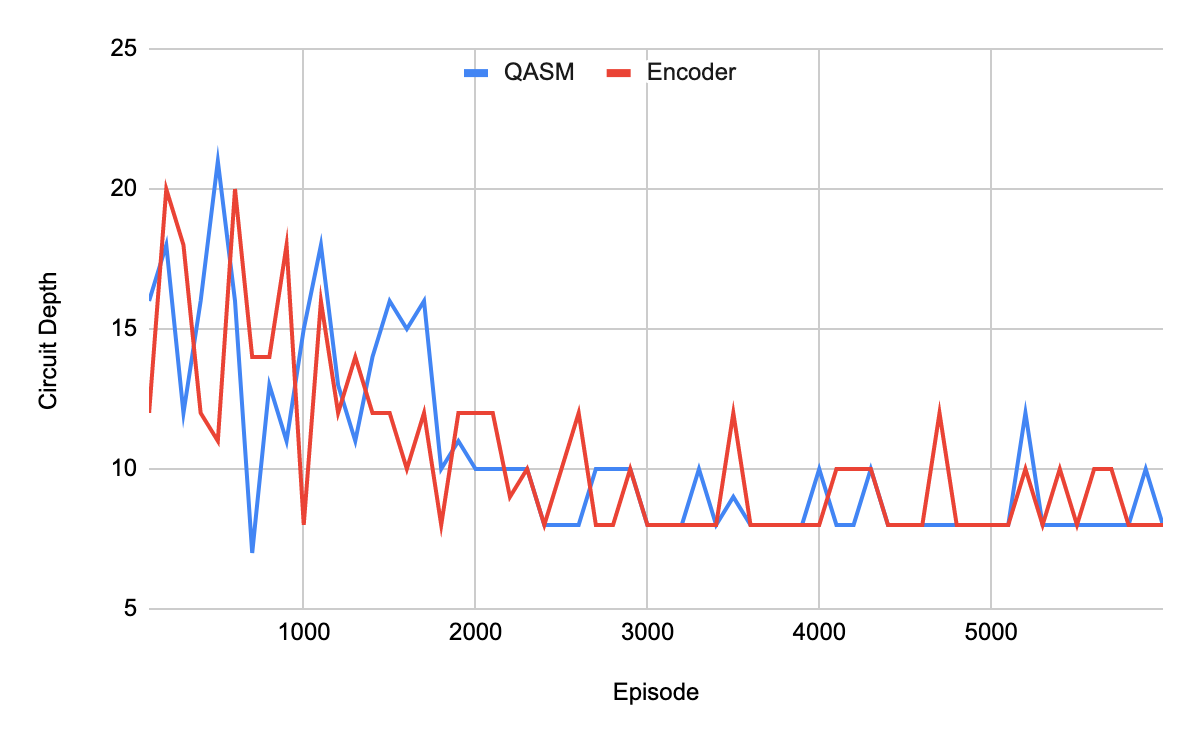}
    \caption{Snapshot of the circuit depth achieved during the training of the RL agent, when using the QASM (blue) and the Encoder (red) representation of circuits. This plot is for Bernstein-Vazirani circuits on 5 qubits where the secret string is $31$. The global optimum would be a circuit of depth 3.}
    \label{fig:prelim}
\end{figure}

The compression of the QTable might influence the effectiveness of the circuit optimisation. For this reason, we compare the quality of the optimised circuits after using both $R_s$ and $R_a$. We conclude that embedding the autoencoder into the RL procedure is effective, because, for our benchmarking circuits, $R_a$, as well as $R_l$, reach the global optimum. Fig.~\ref{fig:prelim} presents an example how the achieved circuit depth evolves during the training of an RL agent when using the novel encoded representation.

\begin{table}[!t]
    \centering
    \begin{tabular}{c|c|c|c|c}
        BV & Epochs & QASM & Encoder & Improv.\\
        \hline
        2 & 3000 & 2535 & 1379 & 45.6\%\\
        3 & 4000 & 7439 & 5131 & 31.02\%\\
        4 & 5000 & 12995 & 10592 & 18.49\%\\
        5 & 6000 & 15880 & 12830 & 19.02\%\\
    \end{tabular}
    \caption{Comparison between the number of RL states when using the character string representation of circuits (\emph{QASM}) and the lossy encoded representation (\emph{Encoder}). The improvement is in the \emph{Improv.} column.}
    \label{tab:compress}
\end{table}

We are evaluating the compression factor achieved when using the autoencoder. Table\ref{tab:compress} contains preliminary results and shows that the mean decrease of the Q-Table size is around 20\% when using the autoencoder.

We conclude, based on our proof-of-concept implementation and from the preliminary data, that our approach achieves a practical compression while not sacrificing the circuit optimization (in Fig.\ref{fig:prelim}) both agents reach the same optimum.

\section{Conclusion}
\label{sec:concl}

We presented a method for scaling the RL optimisation of quantum circuits. We use a graph neural network autoencoder for obtained compressed representations of the circuits encountered during the training of the RL agent. Preliminary results show that the autoencoder compresses by approximately 20\% the number of RL states. The compression does not affect the performance of the RL agent: it reaches the same optimal quantum circuits. Future work will focus on improving the compression, the optimization performance and applying this method to very large scale circuits.

\section*{Acknowledgements}

Ioana Moflic and Alexandru Paler were with funding from the Defense Advanced Research Projects Agency [under the Quantum Benchmarking (QB) program under award no. HR00112230007 and HR001121S0026 contracts]. The views, opinions and/or findings expressed are those of the authors and should not be interpreted as representing the official views or policies of the Department of Defense or the U.S. Government.

\bibliography{__paper}

\end{document}